\documentclass[]{article}
\newcommand{\authorrunning}[1]{}
\usepackage[affil-it]{authblk}
\newcommand{\institute}[1]{\affil{#1}}
\date{}

\usepackage{a4wide}

\usepackage[utf8]{inputenc}
\usepackage[english]{babel}
\usepackage[vlined, ruled]{algorithm2e}
\usepackage{amsmath,amssymb,makeidx,mathrsfs}

\usepackage{amsthm}
\usepackage{xspace}

\usepackage{tikz,pgf}
\usetikzlibrary{shapes}
\usetikzlibrary{arrows}
\usepackage{mathrsfs}
\usepackage{graphicx}
\usepackage{caption}

\usetikzlibrary{decorations.pathmorphing,patterns,scopes}

\newcommand{\ie}{\emph{i.e.}}

\newtheorem{observation}{Observation}

\usepackage[subtle]{savetrees}

\let\oldparagraph\paragraph
\renewcommand{\paragraph}[1]{\oldparagraph{\textbf{#1.}}}
\renewcommand{\paragraph}[1]{\vspace*{0.2cm}\noindent\textbf{#1.}\,\,\,}

\usepackage{layouts} 
\begin{document}

\author{Quentin Bramas}
\authorrunning{Q. Bramas}

\title
{Efficient and Secure TSA for the Tangle}

\institute{ICUBE, Strasbourg University, CNRS, France }

\maketitle


\begin{abstract}
The Tangle is the data structure used to store transactions in the IOTA cryptocurrency. In the Tangle, each block has two parents. As a result, the blocks do not form a chain, but a directed acyclic graph.
In traditional Blockchain, a new block is appended to the heaviest chain in case of fork. In the Tangle, the parent selection is done by the Tip Selection Algorithm (TSA). In this paper, we make some important observations about the security of existing TSAs. We then propose a new TSA that has low complexity and is more secure than previous TSAs.
\end{abstract}

\section{Introduction and Background}
\label{sec:introduction}

A Distributed Ledger Technology (DLT) is a distributed protocol executed by a set of nodes to maintain an append-only data structure. 
In Bitcoin, the data-structure is a chain of blocks, containing transactions. Blocks are appended one after the other to form a chain. Each block requires some amount of computational power, \emph{called weight}, to be created. 
In Bitcoin (and other Proof-of-Work Blockchains), a new block is added to the heaviest branch \ie, the branch that maximizes the sum of the weights of the blocks it contains. This behavior is at the core of the security of Bitcoin.

In this brief announcement, we are interested in the data structure called the \emph{Tangle}, used to store transactions in the IOTA cryptocurrency, and especially in the algorithm used to append new data. We make some important observations about the security of such algorithms and how previous algorithms do not satisfy them. We then propose a new algorithm that is more secure and more efficient than previous solutions.

\paragraph{The Tangle} \emph{The Tangle} is a data-structure where each block of transactions, called \emph{site}, is linked to two previous sites (using hash pointers), called \emph{parents}. The \emph{genesis} site is the only site without parents. Thus, sites form a Directed Acyclic Graph (DAG) of sites. A site is said to \emph{confirm} all its ancestors in the Tangle.  A \emph{tip} of the Tangle is a site which has no child \ie, which is not confirmed by any site.

We consider a network composed of connected nodes that generate and broadcast new sites. Each node has a local copy of the tangle that is updated when a new site is appended.

In order to append a transaction in the Tangle, a node must perform a Proof-of-Work \ie, solving a cryptographic puzzle requiring a certain amount of computational power. The \emph{weight} of a site represents this work and we assume each site has a weight of 1. Then, the \emph{cumulative weight} of a site is defined~\cite{tangle} as the sum of its own weight with the weight of its descendants (the sites that confirm it). 

\paragraph{Tips Selection Algorithm (TSA)}
When a site is added to the Tangle, its parents are selected by a \emph{Tip Selection Algorithm} (TSA). The TSA must select two tips (unconfirmed sites) that are not conflicting (informally, two transactions are conflicting if accepting both would produce a double spend). The TSA is a fundamental component of the protocol because it implicitly indicates how the nodes agree on the current state of the Tangle. Indeed, if two tips are conflicting, the TSA indicates which one is considered correct (and should be extended by appending a new site to it) or orphaned (by ignoring it).

Since each node in the network maintains its own version of the Tangle, a site can end up having multiple children. Indeed, due to the latency in the network, the TSA could chose a site which is a tip locally, but that is already confirmed in another version.
The Tangle whitepaper~\cite{tangle} presents two TSAs\footnote{A third one is briefly presented but is actually just a variation of the MCMC that we present here.}:
\begin{itemize}
    \item Uniform TSA: Each parent is chosen uniformly at random among all the tips.
    \item Markov Chain Monte Carlo (MCMC): the selection of each parent is done by using a random walk. A walker starts from a given site (eg, the genesis), moves from site to child site, and stops when it reaches a tip. The probability of moving to a child site, depends on its cumulative weight (see~\cite{tangle} for more details).
\end{itemize}
\paragraph{The Double Spending Attack}
In the Tangle, an attacker that wants to double spend must generate two conflicting transactions and first broadcast only one of them. When the first transaction is considered well-confirmed (\ie, the honest nodes think the probability to reverse it is small enough), then the attacker can broadcast the second transaction and append a lots of sites (forming \emph{a parasite chain}~\cite{tangle,bramas2018}) so that the first transaction is discarded.

\section{Related Work and Motivations}
\paragraph{Related Work}
The Uniform TSA was initially proposed for its simplicity. One of its advantages is that tips are quickly confirmed~\cite{tangle,kusmierz2019properties}. However, it is easy to see that it offers no protection against double spending attacks. Indeed, an attacker just has to generate more tips than the current number of honest tips to have a higher probability to be selected by honest nodes. Hence, even very old transactions could be canceled easily.

The MCMC algorithm was the first to offer protection against double spending attacks. Indeed, the older a transaction is, the harder it is to cancel it~\cite{tangle,attias2019choose}. However, the MCMC requires computing the cumulative weight of every sites in the tangle (which has worst-case quadratic complexity in the number of sites), and its security depends on a parameter $\alpha$ which also influences the number of tips that are left behind~\cite{kusmierz2019properties} (\ie, tips that are never confirmed). In other words, better security implies less stability, and usability.

\newcommand{\MCMCRW}{MCMC$_{rw}$\xspace}
The efficiency and the security of the MCMC has been improved with \MCMCRW~\cite{attias2019choose}, by using a simpler version of the cumulative weight (which has linear complexity in the number of tips). \MCMCRW obtains a better trade-off security/stability than standard MCMC.
G-IOTA~\cite{g-iota} and E-IOTA~\cite{e-iota} are two extensions of IOTA that proposed mechanisms to limit the number of left-behind tips, while still using MCMC for its security. 

All the previously proposed TSAs mixes in the same algorithm the security and the stability aspects.
Our goal is to give an algorithm that separates these two aspects.

The last version of the IOTA whitepaper~\cite{popov2020coordicide} has a similar approach. It proposes to use a completely distinct algorithm to resolve conflicts so that the TSA is not concerned by the security aspect. However, the security of the proposed consensus algorithm has not yet been formally studied. Our goal is to improve previously defined TSAs, using the same model as the original Tangle whitepaper, which has been formally defined~\cite{tangle,bramas2018}.

\paragraph{Motivations}
Our motivation comes from three important observations.

\begin{observation}
If, between two conflicting transactions, one is considered malicious\footnote{Here malicious just means that it conflicts with a transaction that is considered correct} with higher probability than the other, does it make sense to choose the malicious transaction as parent with non-zero probability ?
\end{observation}
Regardless of the algorithm used to compare conflicting transactions, we believe a transaction that is considered malicious should never be selected as parent, even with small probability. 
Otherwise, a fraction of the honest nodes will support the malicious transactions and help the adversary. 
So, we think a secure TSA should resolves conflicts in a \textbf{deterministic} manner, using another algorithm that we call the \emph{Conflict Resolving Algorithm} (CRA).

\begin{observation}
The uniform random tip selection is the algorithm that offers the best confirmation time and produces the smallest number of tips on average. However it offers poor security guarantees.
\end{observation}
The main reason Uniform random TSA is not used in practice is because it offers poor security guarantees. Indeed, it is very easy for a malicious node to generate a small number of transactions to give a high probability for an old transaction to be selected as parent. However, when there is no conflicts, transactions are confirmed very quickly and no transaction is left over. 
Thus, there is no issue in using the uniform random TSA, after that the set of non-conflicting tips has been deterministically selected.

\begin{observation}
MCMC offers good security guarantees at the price of slower confirmation time and higher number of tips on average.
\end{observation}
Again, if an algorithm provides a good way to discriminate conflicting transactions, then there in no reason not to use it for this purpose. Then, another algorithm can be used to randomly select parents among the non-conflicting remaining tips efficiently. 
The security of MCMC is due to the fact that a random walker has a greater  probability to move towards sites with higher cumulative weight. However, we think there is no need to do it for all the sites, but instead, it should be done only when comparing conflicting sites. 

\section{A New Secure TSA: the two-step TSA}
\label{sec:A New Secure TSA}

\renewcommand{\S}{\mathcal{S}}
\newcommand{\C}{\mathcal{C}}
\paragraph{Model}
Given a Tangle, $\S$ denotes the set of sites. For any subset $C$ of sites, we say that $C$ is \emph{conflict-free} if all the sites in $C$ are pairwise non-conflicting. We now give a more precise definition of tips that takes into account conflicts. We say a conflict-free set $C$ is a \emph{set of tips}, if there is no sites $s\in \S$ and $t\in C$ such that $s$ confirms $t$ and $C\cup\{s\}$ is conflict-free. This means that, if a tip in $C$ is confirmed by some site in $s\in\S$, then $s$ does conflict with another site in $C$. For a site $s$, $w(s)$ denotes its cumulative weight.

\paragraph{The 2-Steps TSA}
Our 2-Step TSA first resolves conflicts between sites and then dispatch parents among conflict-free sites.

Our Conflict Resolver Algorithm (CRA) takes a Tangle and returns a maximal conflict-free set of tips $C$ such that, for any pair of conflicting sites $s_1$ and $s_2$, if $s_1$ is confirmed by some site in $C$, then $w(s_1) \geq w(s_2)$ \ie, the conflict-free set of tips that confirms only the heaviest site in case of conflicts, and is maximal in the sense that no more site can be added to the set without creating conflicts.

\noindent
\begin{minipage}{0.48\textwidth}


\tikzstyle{mirror}=[circle, draw, fill=black,
                        inner sep=0pt, minimum width=6pt]
\tikzstyle{honest}=[circle, draw, fill=white,
                        inner sep=0pt, minimum width=6pt]
\tikzstyle{common}=[circle, draw, fill=white,
                        inner sep=0pt, minimum width=6pt]
\begin{tikzpicture}[scale=0.75]
  \node[opacity=1] at (9.96,9.93) {};
  \node[opacity=1] at (9.96,8) {};

  \path (9.96,7.5) node [common] (v10) {};
  \path (9.96,8.34) node [common] (v12) {};
  \path (10.8,7.32) node [common] (v13) {};

  \draw[->] (v13) -- (v12);
  \draw[->] (v13) -- (v10);        

  \path (11.6,8.8) node [honest, rectangle, minimum height=7pt, line width=1pt] (v8) {};
  \path (12.2,8.56) node [honest] (v14) {};
  \path (13.34,9.18) node [honest] (v15) {};
  \path (14.56,8.72) node [honest] (v16) {};
  \path (14.86,9.34) node [honest] (v17) {};
  \path (13.32,8.54) node [honest] (v18) {};
  \path (12.14,7.66) node [common] (v19) {};
  \path (13.42,7.8) node [honest] (v20) {};
  \path (15.3,8.06) node [honest] (v21) {};
  \path (16.0,8.5) node [honest] (v22) {};
  \path (15.56,9.36) node [honest] (v24) {};
  \path (14.5,8.14) node [honest] (v25) {};
  
  \draw[->] (v14) -- (v13); 
  \draw[->] (v8) -- (v13); 
  \draw[->] (v8) -- (v12); 
  \draw[->] (v15) -- (v14); 
  \draw[->] (v15) -- (v8); 
  \draw[->] (v19) -- (v13); 
  \draw[->] (v14) -- (v8); 
  \draw[->] (v20) -- (v19); 
  \draw[->] (v20) -- (v14); 
  \draw[->] (v18) -- (v19); 
  \draw[->] (v18) -- (v14); 
  \draw[->] (v17) -- (v18); 
  \draw[->] (v17) -- (v15); 
  \draw[->] (v16) -- (v18); 
  \draw[->] (v16) -- (v20);
  \draw[->] (v25) -- (v18); 
  \draw[->] (v25) -- (v20);  
  \draw[->] (v21) -- (v25); 
  \draw[->] (v22) -- (v16); 
  \draw[->] (v22) -- (v21);
  \draw[->] (v24) -- (v17);
  \draw[->] (v24) -- (v16); 
    
  
  \path (12,6.8) node [honest, rectangle, minimum height=7pt, line width=1pt, fill=black!90] (v26) {};
  
  \path (12.5,6.8) node [mirror] (v30) {};
  \path (13,6.8) node [mirror] (v31) {};
  \path (13.5,6.8) node [mirror] (v32) {};
  \path (14,6.8) node [mirror] (v33) {};
  \path (14.5,6.8) node [mirror] (v34) {};
  \path (15,6.8) node [mirror] (v35) {};
  \path (15.5,6.8) node [mirror] (v36) {};
  
  \draw[->, color=black!40] (v26) -- (v13);  
   
  \draw[->] (v30) -- (v26);  
  \draw[->] (v31) -- (v30);  
  \draw[->] (v32) -- (v31);  
  \draw[->] (v33) -- (v32); 
  \draw[->] (v34) -- (v33); 
  \draw[->] (v35) -- (v34); 
  \draw[->] (v36) -- (v35); 
     
  \draw[->, color=black!40] (v31) -- (v13);  
  \draw[->, color=black!40] (v32) -- (v13);  
  \draw[->, color=black!40] (v33) -- (v13); 
  \draw[->, color=black!40] (v34) -- (v13); 
  \draw[->, color=black!40] (v35) -- (v13); 
  \draw[->, color=black!40] (v36) -- (v13);  
    

  \path (15.5,9.8) node [honest] (v41) {};
  \draw[->] (v41) edge[bend right=10] (v12); 

  \path (17,9.5) node [honest, star, star points=7,star point ratio=0.55] (v40) {};
  \draw[->] (v40) edge (v24);  
  \draw[->] (v40) edge[bend right=10] (v41);  

\draw [dashed] plot [smooth cycle] coordinates {(15.2,9.8) (15.9,9.8) (16.4,7.8) (15.85,7.8)};
\node[text opacity=1, fill=white, inner sep = 0] at (16.3,7.8) {$C$};

\end{tikzpicture}

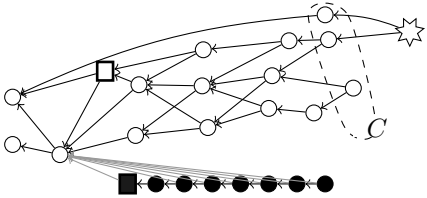
\captionof{figure}{In this example, the white square is considered correct and the black one is discarded.}
\label{fig:adversarial extension of G}
\end{minipage}\hfill
\begin{minipage}{0.48\textwidth}

\tikzstyle{mirror}=[circle, draw, fill=black,
                        inner sep=0pt, minimum width=6pt]
\tikzstyle{honest}=[circle, draw, fill=white,
                        inner sep=0pt, minimum width=6pt]
\tikzstyle{common}=[circle, draw, fill=white,
                        inner sep=0pt, minimum width=6pt]
\begin{tikzpicture}[scale=0.75]

  \path (9.96,7.5) node [common] (v10) {};
  \path (9.96,8.34) node [common] (v12) {};
  \path (10.8,7.32) node [common] (v13) {};

  \draw[->] (v13) -- (v12);
  \draw[->] (v13) -- (v10);        

  \path (11.6,8.8) node [honest] (v8) {};
  \path (12.2,8.56) node [honest] (v14) {};
  \path (13.34,9.18) node [honest] (v15) {};
  \path (14.56,8.72) node [honest] (v16) {};
  \path (14.86,9.34) node [honest] (v17) {};
  \path (13.32,8.54) node [honest] (v18) {};
  \path (12.14,7.66) node [common] (v19) {};
  \path (13.42,7.8) node [honest] (v20) {};
  \path (15.3,8.06) node [honest] (v21) {};
  \path (16.02,8.7) node [honest] (v22) {};
  \path (17.22,9.02) node [honest, rectangle, minimum height=7pt, line width=1pt, fill=black!90] (v23) {};
  \path (15.96,9.36) node [honest] (v24) {};
  \path (14.5,8.14) node [honest] (v25) {};
  
  \draw[->] (v14) -- (v13); 
  \draw[->] (v8) -- (v13); 
  \draw[->] (v8) -- (v12); 
  \draw[->] (v15) -- (v14); 
  \draw[->] (v15) -- (v8); 
  \draw[->] (v19) -- (v13); 
  \draw[->] (v14) -- (v8); 
  \draw[->] (v20) -- (v19); 
  \draw[->] (v20) -- (v14); 
  \draw[->] (v18) -- (v19); 
  \draw[->] (v18) -- (v14); 
  \draw[->] (v17) -- (v18); 
  \draw[->] (v17) -- (v15); 
  \draw[->] (v16) -- (v18); 
  \draw[->] (v16) -- (v20);
  \draw[->] (v25) -- (v18); 
  \draw[->] (v25) -- (v20);  
  \draw[->] (v21) -- (v25); 
  \draw[->] (v22) -- (v16); 
  \draw[->] (v22) -- (v21);
  \draw[->] (v24) -- (v17);
  \draw[->] (v24) -- (v16); 
  \draw[->] (v23) -- (v22); 
  \draw[->] (v23) -- (v24);  
    
  
  \path (12,6.8) node [honest, rectangle, minimum height=7pt, line width=1pt] (v26) {};
  
  \path (12.5,6.8) node [honest] (v30) {};
  \path (13,6.8) node [honest] (v31) {};
  \path (13.5,6.8) node [honest] (v32) {};
  \path (14,6.8) node [honest] (v33) {};
  \path (14.5,6.8) node [honest] (v34) {};
  \path (15,6.8) node [honest] (v35) {};
  \path (15.5,6.8) node [honest] (v36) {};
  
  \draw[->, color=black!40] (v26) -- (v13);  
   
  \draw[->] (v30) -- (v26);  
  \draw[->] (v31) -- (v30);  
  \draw[->] (v32) -- (v31);  
  \draw[->] (v33) -- (v32); 
  \draw[->] (v34) -- (v33); 
  \draw[->] (v35) -- (v34); 
  \draw[->] (v36) -- (v35); 
     
  \draw[->, color=black!40] (v31) -- (v13);  
  \draw[->, color=black!40] (v32) -- (v13);  
  \draw[->, color=black!40] (v33) -- (v13); 
  \draw[->, color=black!40] (v34) -- (v13); 
  \draw[->, color=black!40] (v35) -- (v13); 
  \draw[->, color=black!40] (v36) -- (v13);  
    

  \path (17,7.8) node [honest, star, star points=7,star point ratio=0.55] (v40) {};
  \draw[->] (v40) -- (v36);  
  \draw[->] (v40) -- (v22);  

\draw [dashed] plot [smooth cycle] coordinates {(15.8,9.8) (16.3,9.5) (16.3,7.5) (15.6,6.5) (15.2,6.5) (15.7,7.8)};
\node[text opacity=1, fill=white, inner sep = 0] at (15.7,7.6) {$C$};

\end{tikzpicture}

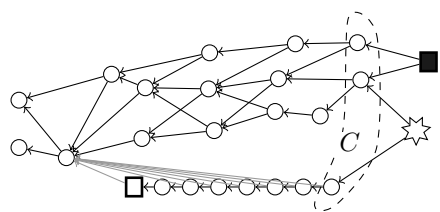
\captionof{figure}{In this example, the new site can merge both branches.}
\label{fig:adversarial extension of G 2}
\end{minipage}

Our Tip Dispatcher Algorithm (TDA) takes a set of conflict-free tips $C$ and returns two tips $p_1$ and $p_2$ selected uniformly at random among $C$, with $p_1$ and $p_2$ distinct if $|C| \geq 2$.

In Figure~\ref{fig:adversarial extension of G} we see a tangle and two conflicting transactions (the two squares). Our CRA first discriminates between the two and considers the white square to be correct and discards the black sites (all the sites confirming the black square). The output of the CRA is the set $C$ containing three conflict-free tips. Then our TDA dispatches the two parents without discriminating between the old and the recent sites. The goal of the TDA is to confirm as many sites as possible, reducing the number of left-over sites.

\paragraph{Security} 
Using our algorithm, if the honest nodes agree on a conflict-free set of tips, then they all extend the tangle in the same way, increasing the weight of the same set of sites. In other words, for any discarded site, there is a site, considered correct, whose weight increases for each new honest site.

This implies that, if an adversary wants to discard a site that is considered correct, it has to generate sites at a higher rate than the honest nodes (which is a necessary assumption anyway~\cite{bramas2018}).
It means that, like in Bitcoin, the probability of creating a successful double spending attack on a site decreases exponentially fast with its weight.

This property is not obtained by previous TSAs. For instance, if a parasite chain has probability $1/3$ to be selected by the MCMC, then an honest node will append one third of its transactions in the parasite chain (assuming they are independent), which is not the intended behavior. In addition, $1/3$ of the honest transactions globally will end up selecting the parasite chain as the correct one. A third of the honest nodes plus the malicious node then represents half of the computational power, so that it becomes even easier for the malicious node to increase the probability of selecting its parasite chain.

Using our TSA, the parasite chain is never selected if its cumulative weight is smaller than another branch of the tangle. This implies that a malicious node that wants to double spend has to create a parasite chain on its own and is not helped by honest nodes. 

Another interesting property of our TSA is that it does not automatically consider correct a site that is located on the main branch. Instead, it compares conflicting sites independently on where they are located on the tangle. Doing so, we can confirm a separate branch that could look like a parasite chain, but is in fact older and might contain honest transactions as well, for instance if it was generated offline. Indeed, we do not want to discard an entire chain just because a conflicting site appears on top of the main chain. Figure~\ref{fig:adversarial extension of G 2} illustrates the situation. We see that the white square has a greater cumulative weight compared to the black square, so only the site confirming the black square (there is no such site in this example) are discarded, creating two tips (the parents of the black-square site). We then have a chance to merge the two branches with a new site (the star-shaped one) using our TDA.

In this situation the MCMC would choose the main branch with greater probability and would almost never merge both branches since the MCMC would never stops its random walk to a parent of black square because it is not a tip.

%

\paragraph{Performances} 
Despite using the cumulative weight, which is computed in $\Theta(n)$ time for a given site, our algorithm can have constant complexity in most situations.

After receiving the Tangle from its peers, a node can compute the conflict-free set of tips $C$ with the CRA, while storing the cumulative weight of each site for later use. After that, every time the node has to generate a site $s$, the TSA will return two parents $p_1$ and $p_2$ among $C$ and there is no need to run the CRA again for the next site as the new conflict-free set of tips is simply $C\cup\{s\}\setminus\{p_1, p_2\}$.
Similarly, for each incoming site $s$, if $s$ confirms a site in $C$, then we know $s$ is considered correct and we can update $C$ by adding $s$ and removing the confirmed tips. So if all the nodes are honest, after the first run of the CRA, every execution of the TSA has constant-time complexity.

However, if an incoming site confirms a site $s_{m}$, considered malicious, and conflicting a site $s_c$ considered correct, then we can increment the weight of $s_m$ by one and compare it to the previously computed weight of $s_c$. If the weight of $s_m$ is still smaller than the weight of $s_c$, we can safely ignore the new site as running the CRA again will not change our current conflict-free set of sites $C$. If the weight of $s_m$ becomes greater than the previously computed weight of $s_c$, then we have to update the weight of $s_c$ and do the comparison again. We believe other optimizations could be performed in this case as well.

\paragraph{Concluding Remarks} We propose a new paradigm for constructing secure and efficient TSAs. We observed that existing TSAs can be improved by spliting the parent selection into a conflict resolving phase and tip dispatcher phase. We believe this work will open new research on the security and the performances of TSAs.

\bibliographystyle{splncs04}
\bibliography{biblio}
\label{sec:biblio}

\end{document}